# Scientific Computing in the Cloud

J. J. Rehr, J. P. Gardner, M. Prange, L. Svec and F. Vila

Department of Physics, University of Washington, Seattle, WA 98195

(December 30, 2008)

ABSTRACT

We investigate the feasibility of high performance scientific computation using *cloud computers* as an alternative to traditional computational tools. The availability of these large, virtualized pools of compute resources raises the possibility of a new compute paradigm for scientific research with many advantages. For research groups, cloud computing provides convenient access to reliable, high performance clusters and storage, without the need to purchase and maintain sophisticated hardware. For developers, virtualization allows scientific codes to be optimized and pre-installed on machine images, facilitating control over the computational environment. Preliminary tests are presented for serial and parallelized versions of the widely used x-ray spectroscopy and electronic structure code FEFF on the Amazon Elastic Compute Cloud, including CPU and network performance.

## I.  Introduction

Amazon Web Services (AWS) [1] is an important provider of "Cloud Computing" (CC) services which are in wide use. However, these virtual computational tools have scarcely been exploited for scientific computing applications, which is the aim of this investigation. Of primary interest to computational scientists are Amazon's Elastic Compute Cloud (EC2) environment [2] and Simple Storage Service (S3) [3]. The goal of this project is to push the limits of the CC paradigm in terms of both usability and performance in order to evaluate how beneficial these services can be for high performance scientific computing (HPC). Our work thus far has focused on porting the FEFF code [4] to the Cloud. This widely used scientific code calculates both the x-ray spectra and electronic structure of complex systems in real-space on large clusters of atoms, and runs on a variety of computing environments. Here we present preliminary results that benchmark the performance of both serial and parallel versions of FEFF on EC2 virtual machines. In addition we have developed a set of tools that enable a user to deploy their own virtual compute clusters, both for FEFF and other parallel codes.

## II.  Achievements

Our recent work is divided in two main areas: i) *Benchmarking*, where we have tested the performance of the FEFF codes on EC2 hardware, and ii) *Development*, where we have started to create an environment which permits the FEFF user community to run different versions of the software in their own EC2-resident compute clusters. Both of these efforts required that we first gain an understanding of the AWS infrastructure.

## A. Understanding the EC2 and S3 infrastructures

The Amazon EC2 is a service that hosts "Amazon Machine Images" (AMIs) on generic hardware located "somewhere" within the Amazon computer network. Amazon offers a set of public AMIs that users can take and customize to their needs. Once an AMI has been modified, users can store it in their Amazon S3 accounts for subsequent reuse later. The "elasticity" of EC2 denotes the ability of a user to spawn an arbitrary number of "instances" of AMIs, while scaling the computational resources to match instantaneous computational demands as needed.

The EC2 and S3 provide three sets of tools for creating and using AMIs. The AMI tools [5] are command-line utilities used to bundle the current state of an image and upload it to S3 storage. The API tools [6] serve as the client interface to the EC2 services. They allow the user to register, launch, monitor and terminate AMI instances. Finally, S3 provides a set of libraries [7] that allow the developer to interact with the S3 server to manage stored files such as AMIs. These tools are available in several formats. Currently we use the Python, Ruby and Java implementations, supplemented with our own set of Bash and Python scripts. This variety of coding languages gives developers a wide range of options, allowing them to tailor their implementations to optimize results. In order for *developers* to use AWS, all three sets of tools are needed. On the other hand, *users* only need the API tools, unless they wish to modify and store our pre-configured images.

We have also experimented with Elasticfox [8] and S3fox [9], two graphical user interfaces that provide partial implementations of the EC2 and S3 tools. These are extensions of the Firefox browser that provide an excellent alternative to the above tools for new users of the AWS services. Elasticfox gives an all-in-one picture of the current state of the users' AMIs and of the instances that are currently active. It also allows the user to initiate, monitor and terminate AMI instances. S3fox mimics the interface of many commonly used file transfer programs, allowing the user to create and delete storage areas in S3. We believe these graphical browser extensions will prove very useful and intuitive for the majority of our user community. Therefore, we plan to develop our user environment so that these tools can be used in conjunction with it.

## B. Implementation of FEFF and JFEFF on EC2

Using a public AWS AMI with Fedora 8, we have created a FEFF AMI which provides both the command line driven FEFF code (version 8.4, denoted below as FEFF84) and JFEFF, a graphical user interface (GUI) which facilitates the execution of FEFF84. In order to get these programs running, we enhanced the template AMI with X11 and a JAVA run-time environment. JFEFF, the JAVA-based FEFF84 GUI, then functions as if it were running locally. Figure 1 shows a screenshot of JFEFF running on EC2, the FEFF AMI console, and in the background the Elasticfox control screen. As might be expected, the only notable limitation observed is the relatively slow response of the GUI when running over a network.

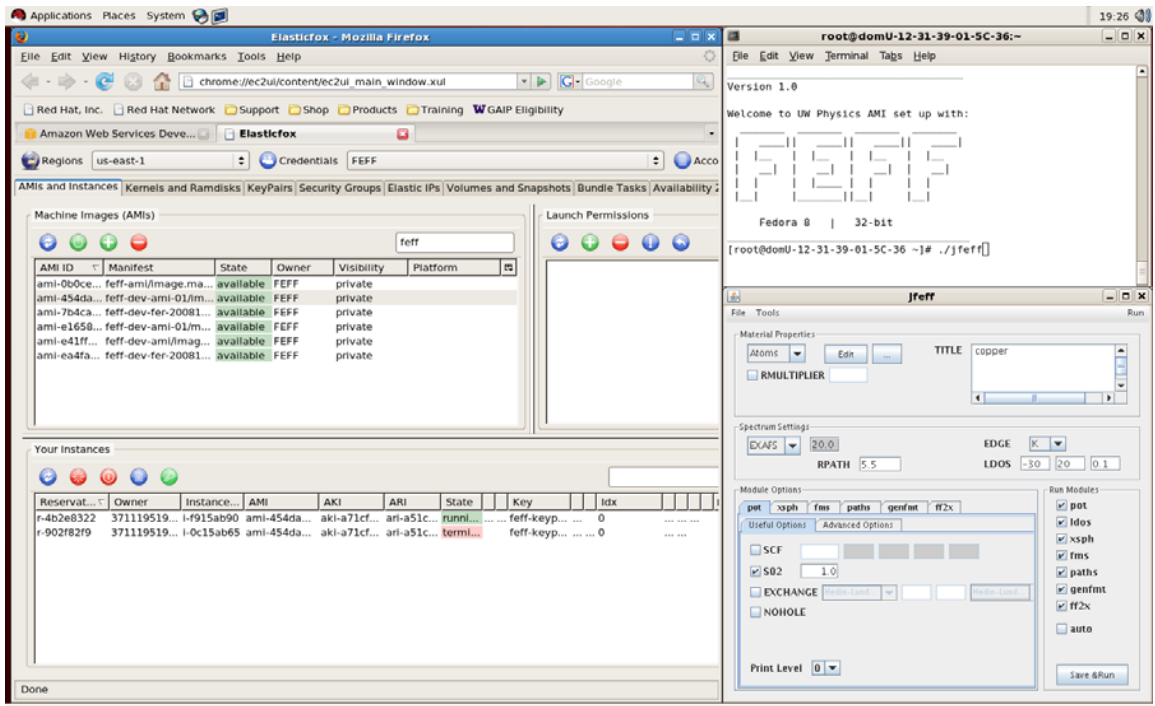

**Figure 1:** Screenshot of FEFF EC2 console (upper right), JFEFF (lower right) and in the background the Elasticfox browser screen.

## C. Testing FEFF84 serial performance

As part of our overarching goal of high performance scientific computing on the Cloud, our FEFF AMI was first used to test the serial performance of FEFF84 on instances with different computing power. Figure 2 shows runtime *vs* atomic cluster size results for a typical full multiple scattering FEFF calculations for Boron Nitride, including atomic cluster sizes ranging from 29 to 157 atoms; typically about 100 atoms is sufficient to obtain converged spectra. Two instance types, both running 32-bit operating systems, were used: i) a "small" instance using a 2.6 GHz AMD Opteron processor and ii) a "medium" instance using a 2.33 GHz Intel Xeon processor, both compiled with gfortran. For comparison we have included results from one of our local systems using a 64-bit 2.0 GHz AMD Athlon processor. Also shown in Figure 2 are the results obtained with a highly optimized version of FEFF84. The compiled executable was produced on our HPC AMD Opteron cluster in the Department of Physics at UW using the PGI Fortran compiler. This compiler produces very efficient code on AMD processors. As can be seen in Figure 2, the resulting code makes the small instance even faster than the medium one. Consequently, we believe that well optimized HPC tools should be included in the AMIs to provide good performance for scientific applications.

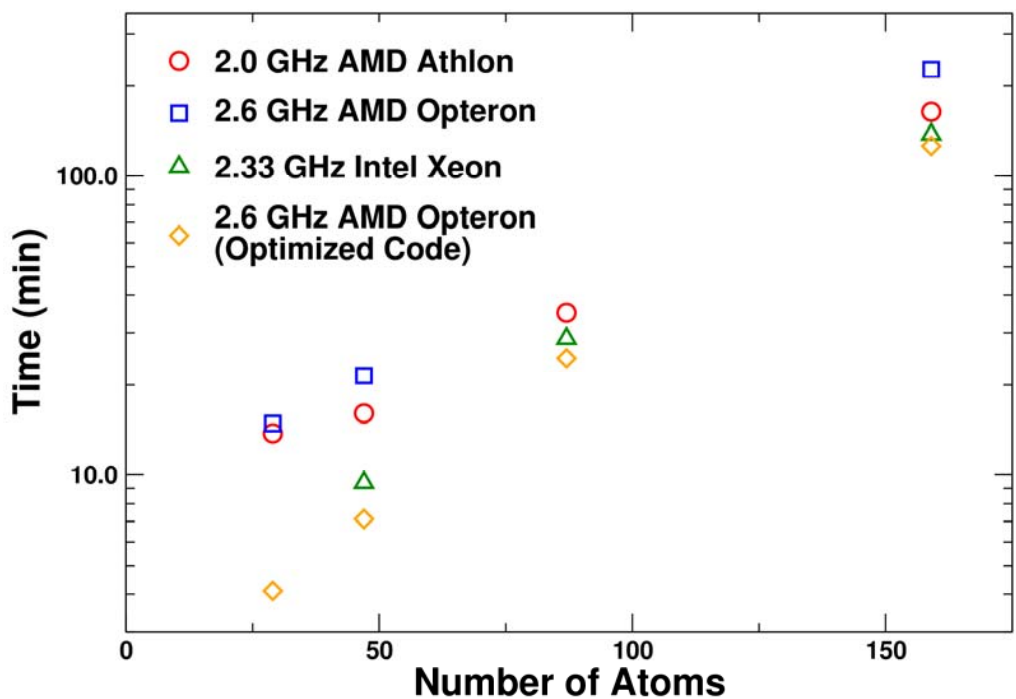

**Figure 2: Comparison of the serial performance (run-time *vs* cluster size) of different AMI instance types for typical FEFF84 calculations.**

### D. Development of strategies and tools for parallel cluster computing

The FEFF user community includes a growing number of users with parallel computation experience who are familiar with Linux and MPI, and others who are virtually helpless in such HPC environments. While the HPC versions of FEFF are becoming increasingly important for studies of complex materials, many users lack access to adequate HPC resources. One of the advantages of Cloud Computing is the potential to provide such access without the need to purchase and maintain or even understand HPC hardware.

Therefore, we have adopted a two-pronged development strategy that serves both kinds of users: i) We provide a set of generic Python tools that can be used to launch any suitably configured Amazon EC2 Linux image for parallel computation. This will give the user who is accustomed to using MPI on the workstation or cluster an immediately useful way to run both our FEFF MPI software and possibly any other MPI parallel code on EC2. Since Python is widely used in the scientific computing community, many users should also be able to customize these scripts to their needs. ii) Our second approach provides a complete, standalone FEFF MPI parallel runtime environment, where users need only to know their AWS account credentials but otherwise do not require any experience with parallel codes, hardware, or the EC2. We believe a successful development along these lines could be greatly beneficial to the scientific community.

As discussed above, Amazon Web Services provides a set of Python modules designed to interact with EC2. Based on these modules, other developers have produced a set of Python scripts for managing MPI clusters on EC2 [10]. However, these scripts are relatively primitive at present, and have many limitations. Moreover, changes in Amazon's API reporting format has rendered some of these scripts unusable. Nevertheless, we have been able to build on these scripts to create four improved, easy-to-use Python scripts for managing an MPI cluster on EC2:

| | |
|---|---|
| `ec2-cluster-start.py` | Starts a cluster of a given number of nodes. |
| `ec2-cluster-status.py` | Reports the status of the instances in the cluster (e.g. "pending", "running", etc). |
| `ec-cluster-config.py` | Provisions every node with the required MPI configuration information and necessary SSH keys, once all the instances are running. |
| `ec-cluster-stop.py` | Terminates the MPI cluster instances. |

All of these scripts are capable of managing multiple clusters simultaneously. When users start a cluster, they can optionally give it a name (e.g. "cluster1"). The scripts then can associate this name with the proper EC2 reservation ID and instances. In this way, users can easily spawn an arbitrary number of EC2 clusters, each with a different name, and manage them easily. To run an MPI job, the user simply logs into the MPI master node and executes a given "mpirun" task of their choice. These scripts are generic in that they contain nothing specific to FEFF and can therefore be used to manage any set of AMIs that have a correct version of MPI installed and a non-root user account. Thus, these tools should be of use to any scientist who wishes to use EC2 for HPC. We therefore plan to release them to the scientific community, e.g., on the FEFF WWW site [4].

In addition to these scripts, we are developing a set of original tools tailored to the FEFF AMI and aimed at typical FEFF MPI users interested in HPC calculations on complex systems. They consist of three Bash scripts that i) launch a cluster, ii) allow the user to connect to it, and iii) to terminate it: i) The `cluster_launch` script boots the required number of instances and monitors them until all are available. Then it gathers information about the instances (private and public instance names, IP addresses, etc.) and creates the configuration files needed to use an MPI cluster efficiently. This includes a list of hosts in the cluster and a setup to share information between the images. The initialization is finalized with the transfer of SSH keys needed to access a user specifically configured to run FEFF. ii) The user can then login using the `cluster_connect` script, which transparently takes him/her to the appropriate user on the head instance. iii) Finally, the cluster instances are terminated with the `cluster_terminate` script. These tools have allowed us to turn a conventional laptop into a computer that controls a virtual supercomputer. It should be noted that although these scripts were designed to work with the FEFF AMI, many MPI programs share a run infrastructure similar to that of FEFF, and hence could be easily modified to use it.

### E. Testing FEFF84 parallel performance

The tools described in the previous section have allowed us to perform some preliminary studies of scalability of the parallel version of FEFF84 (FEFF84MPI) on EC2. Figure 3 shows a comparison of the scaling *vs* number of CPUs observed in EC2 and that obtained on our local 1.8 GHz AMD Opteron HPC cluster. The similarity of the results indicates that the use of a virtual EC2 cluster does not degrade the parallel performance of FEFF84 compared to that on a conventional physical computer cluster.

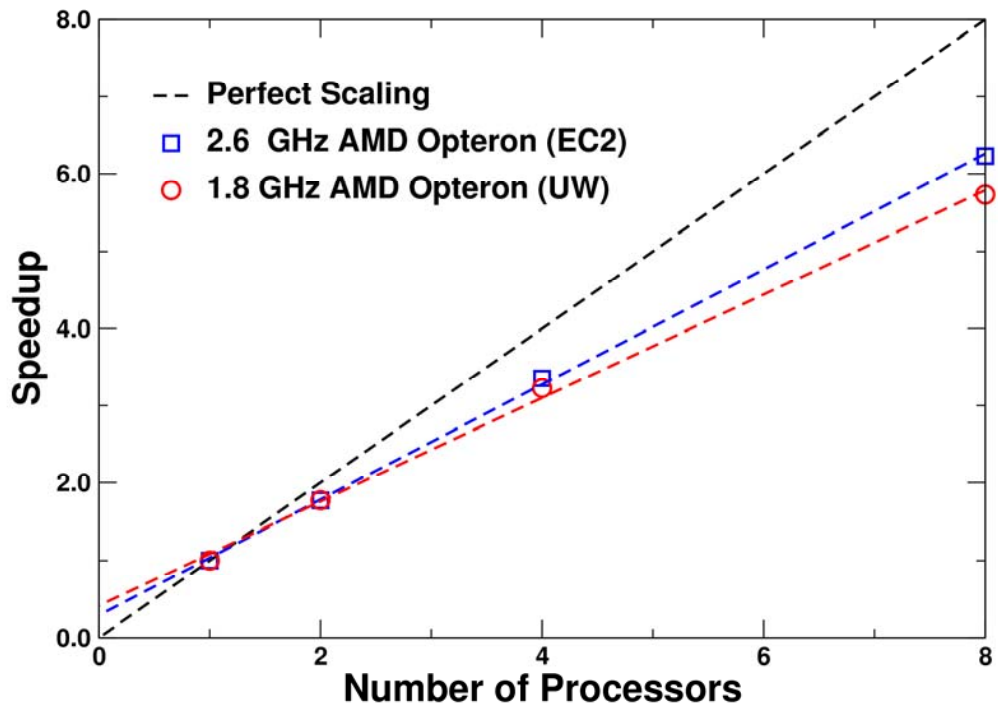

**Figure 3: Comparison of the scalability of FEFF84MPI observed for the EC2 to that observed on our local 1.8 GHz Opteron cluster**

## III. Remaining challenges and goals

In the near future we plan to address several remaining issues. Of special interest is the development of high performance computing AMIs. These images would use 64-bit operating systems and include high performance compilers and numerical libraries. We believe these images will provide us with a representative production environment to assess the overall economic feasibility of high performance scientific cloud computing. Our automatic cluster tools will also allow us to further test and optimize interprocess communication. As demonstrated above, this is likely a minor issue for a moderately coupled program like FEFF, but could be very important for many other commonly-used codes. We plan to modify our scripts to run optionally from a special "master" AMI. Within this setup, users would not have to download our scripts to their workstations.

Instead, they would simply request a "master" instance, login, and manage their MPI clusters from it. We will also develop easy-to-use tools to transfer files both to and from the head instance. Finally, we will explore the possibility of creating a unified front-end, similar to Elasticfox for the management of MPI clusters on the EC2. If successful, we plan to incorporate this feature into the JFEFF GUI to FEFF. Such and approach would ensure that the GUI is not hindered by the network performance, increase the user-friendliness of running FEFF with MPI, and reduce the size of the FEFF AMI.

Although this report is only preliminary, we already find the AWS EC2 to be reasonably adaptive to the goal of making high performance scientific computation readily available to the scientific community. We have already explored a variety of promising methods which permit users to interact with our images. These range from predefined scripts that manage our AMIs locally to tools which allow users to remotely SSH into our AMIs and control images directly. A further account of these developments will be presented at the March 2009 Meeting of the American Physical Society [11].

**Acknowledgments:** This work is primarily supported by the National Science Foundation through NSF grant DMR-0848950, and in part by NSF CDI grant PHY-0835543 (JG and FV). The FEFF project is supported primarily through DOE grant DE-FG03-97ER45623. We especially thank Amazon.com personnel including J. Barr, T. Laxdal, P. Sirota, P. Sridharan, R. Valdez, and W. Vogels, for their enthusiastic support of this project and for providing AWS resources. Some of the tools developed in this project are based on the MPI Cluster Tools of Peter Skomoroch at Datawrangling.

## References
[1] http://aws.amazon.com/
[2] http://aws.amazon.com/ec2/
[3] http://aws.amazon.com/s3/
[4] http://leonardo.phys.washington.edu/feff/; see also *Real space multiple scattering calculation and interpretation of X-ray absorption near edge structure*, A. Ankudinov, B. Ravel, J.J. Rehr, and S. Conradson, Phys. Rev. B **58**, 7565 (1998), and *Parallel calculation of electron multiple scattering using Lanczos algorithms*", A. L. Ankudinov, C. E. Bouldin, J. J. Rehr, J. Sims, and H. Hung, Phys. Rev. B **65**, 104107 (2002).
[5] http://developer.amazonwebservices.com/connect/entry.jspa?externalID=368
[6] http://developer.amazonwebservices.com/connect/entry.jspa?externalID=351
[7] http://developer.amazonwebservices.com/connect/kbcategory.jspa?categoryID=47
[8] http://developer.amazonwebservices.com/connect/entry.jspa?externalID=609
[9] http://developer.amazonwebservices.com/connect/entry.jspa?externalID=366
[10] http://www.datawrangling.com/mpi-cluster-with-python-and-amazon-ec2-part-2-of-3
[11] http://meetings.aps.org/Meeting/MAR09, J. J. Rehr et al., T13.00008.